\documentclass[aps,pre,twocolumn,groupedaddress]{revtex4-2}
\usepackage{bm}
\usepackage{dcolumn}
\usepackage{braket}
\usepackage{graphicx}
\usepackage{amssymb}
\usepackage{subcaption}

\begin{document}
\title{Ising formulation of integer optimization problems for utilizing quantum annealing\\ in iterative improvement strategy}

\author{Shuntaro Okada$^{1,2}$}
\author{Masayuki Ohzeki$^{2,3,4,5}$}
\affiliation{$^1$AI R{\&}I Division, DENSO CORPORATION, Aichi, Japan}
\affiliation{$^2$Graduate School of Information Sciences, Tohoku University, Sendai, Japan}
\affiliation{$^3$International Research Frontier Initiative, Tokyo Institute of Technology, Tokyo, Japan}
\affiliation{$^4$Department of Physics, Tokyo Institute of Technology, Tokyo, Japan}
\affiliation{$^5$Sigma-i Co., Ltd., Tokyo, Japan}

\date{\today}

\begin{abstract}
Quantum annealing is a heuristic algorithm for searching the ground state of an Ising model.
Heuristic algorithms aim to obtain near-optimal solutions with a reasonable computation time. Accordingly, many algorithms have so far been proposed.
In general, the performance of heuristic algorithms strongly depends on the instance of the combinatorial optimization problem to be solved because they escape the local minima in different ways.
Therefore, combining several algorithms to exploit their complementary strength is effective for obtaining highly accurate solutions for a wide range of combinatorial optimization problems.
However, quantum annealing cannot be used to improve a candidate solution obtained by other algorithms because it starts from an initial state where all spin configurations are found with a uniform probability.
In this study, we propose an Ising formulation of integer optimization problems to utilize quantum annealing in the iterative improvement strategy.
Our formulation exploits the biased sampling of degenerated ground states in transverse magnetic field quantum annealing.
We also analytically show that a first-order phase transition is successfully avoided for a fully connected ferromagnetic Potts model if the overlap between a ground state and a candidate solution exceeds a threshold.
The proposed formulation is applicable to a wide range of integer optimization problems and enables us to hybridize quantum annealing with other optimization algorithms.
\end{abstract}
\maketitle

\section{Introduction}
The minimization problem of a cost function defined by discrete variables is called a combinatorial optimization problem, and it has many real-world applications.
In general, a combinatorial optimization problem can be mapped to the ground state search of an Ising model \cite{NP_statistical_mechanics, Ising_mapping}.
Therefore, a wide range of combinatorial optimization problems are expected to be efficiently solved if an Ising-type computer that quickly identifies the ground state of Ising models is achieved.
Several Ising-type computers, such as the D-Wave quantum annealer \cite{D-wave_machine} and the Fujitsu digital annealer \cite{digital_annealer}, have been actively developed to construct a general solver of combinatorial optimization problems.
The D-Wave quantum annealer, which achieves the hardware implementation of quantum annealing (QA) \cite{QA_original}, has attracted great attention in recent years.
Many studies have demonstrated the applicability of the D-Wave quantum annealer to practical problems \cite{D-Wave_application1, D-Wave_application2, D-Wave_application3, D-Wave_application4, D-Wave_application5, D-Wave_application6, D-Wave_application7, D-Wave_application8, D-Wave_application9, D-Wave_application10, D-Wave_application11, D-Wave_application12, D-Wave_application13, D-Wave_application14, D-Wave_application15, D-Wave_application16, D-Wave_application17, D-Wave_application18, D-Wave_application19, D-Wave_application20, D-Wave_application21, D-Wave_application22}.

Inspired by simulated annealing (SA) \cite{SA_original}, QA was originally proposed as a method for searching the ground state of Ising models.
QA and SA are heuristic algorithms that aim to obtain near-optimal solutions with a reasonable computation time.
If the landscape of the cost function is complicated and has many local minima, it is important to escape the local minima and search for as many local minima as possible to obtain highly accurate solutions.
SA employs thermal fluctuations to escape the local minima, while QA escapes the local minima through the tunneling effects induced by quantum fluctuations.
A quantum fluctuation is usually induced by a transverse magnetic field.
Strong fluctuations are applied at the beginning of QA and SA. All spin configurations are found with a uniform probability at the initial state.
Subsequently, the fluctuations are gradually decreased to narrow down the phase space range to be searched. A candidate solution, whose energy is hopefully optimal, is then finally obtained.
The probability of obtaining a ground state becomes higher as the system is changed slower \cite{SA_convergence1, SA_convergence2, adiabatic_theorem}.
The adiabatic theorem \cite{adiabatic_theorem} states that the computation time of QA is proportional to the inverse square of the minimum energy gap between the instantaneous ground state and the first excited state.
Although numerous studies have compared the performances of QA and SA \cite{QA_SA_compare1, QA_SA_compare2, QA_SA_compare3, QA_SA_compare4, QA_SA_compare5, QA_SA_compare6, D-wave_compare1, D-wave_compare2, D-wave_compare3}, the superiority of QA to SA remains elusive.

One of the main strategies of heuristic algorithms is the iterative improvement of a candidate solution.
In addition to QA and SA, hill climbing, tabu search \cite{tabu1, tabu2}, and genetic algorithm are also famous examples.
The abovementioned algorithms escape the local minima in different ways; hence, the performance of heuristic algorithms generally depends on the instance of the combinatorial optimization problem to be solved.
Combining several algorithms to exploit their complementary strength is effective in obtaining highly accurate solutions for a wide range of combinatorial optimization problems.
For example, a candidate solution obtained by the genetic algorithm is next improved by the tabu search.
However, QA and SA cannot be used to improve a candidate solution in its original form because the probability of obtaining a spin configuration is uniformly distributed in the initial state.
To resolve this issue, reverse annealing (RA) \cite{reverse_annealing1, reverse_annealing2, reverse_annealing3, reverse_annealing4} is proposed as a refined QA that starts from a candidate solution.
The Hamiltonian of RA consists of three terms: problem, driver, and initial Hamiltonian.
Here, the problem Hamiltonian represents the Ising formulation of the cost function to be minimized. The driver Hamiltonian induces the quantum fluctuations. The ground state of the initial Hamiltonian is identical to the candidate solution.
The Hamiltonian of RA is set to the initial Hamiltonian at the beginning of RA.
Next, RA interpolates the initial and problem Hamiltonian, during which the quantum fluctuation strength is first increased, and then decreased to zero at the end.
The search space of RA is restricted around the candidate solution by introducing an extra energy term, namely the initial Hamiltonian.
A coefficient of the initial Hamiltonian must then be changed in conjunction with that of the driver Hamiltonian.
Analytically, a first-order phase transition in the $p$-spin model can be avoided if the overlap between the candidate solution and the ground state exceeds a threshold and if the coefficients of the initial and driver Hamiltonian are appropriately controlled \cite{ reverse_annealing2}.

In this study, we propose an Ising formulation of integer optimization problems to use QA to improve a candidate solution.
Information on the candidate solution is embedded into the problem Hamiltonian by exploiting the biased sampling of degenerated ground states in transverse magnetic field QA (TMF-QA) \cite{unfair_sampling1, unfair_sampling2, unfair_sampling3, unfair_sampling4, unfair_sampling5}.
Our method does not introduce an extra term to the Hamiltonian of QA. In addition, the number of control parameters does not increase from the original form.
The proposed formulation is applied to a fully connected ferromagnetic (FC-FM) Potts model.
It is analytically shown that a first-order phase transition is successfully avoided if the overlap between the candidate solution and the ground state exceeds a threshold.
The minimum energy gap in a system with a first-order phase transition typically decreases exponentially with the system size \cite{first-order_exponential1, first-order_exponential2, first-order_exponential3},
while that in a system with a second-order phase transition polynomially decreases.
Therefore, using the proposed formulation, the computation time of QA is expected to be significantly improved if a near-optimal solution is prepared a priori.
The proposed formulation can be applied to a wide range of integer optimization problems and enables us to apply QA to the improvement of a candidate solution obtained by other optimization algorithms.

The remainder of this paper is organized as follows:
Section II briefly explains a conventional Ising formulation of integer optimization problems
and presents the proposed Ising formulation to embed information about the candidate solution into the problem Hamiltonian;
Section III explains the application of the proposed formulation to TMF-QA of the FC-FM Potts model and provides an analytical investigation of the phase transition order for various overlap values between the candidate solution and the ground state; and
finally, Section IV presents the discussion and conclusions of this study.

\section{Ising formulation of integer optimization problems}
After a brief explanation of a conventional formulation, we will propose herein the Ising formulation of integer optimization problems for utilizing QA to improve a candidate solution.

\subsection{Conventional formulation}
In solving integer optimization problems using Ising-type computers, the cost function must be formulated as a Hamiltonian of the Ising model.
Several formulations have so far been proposed\cite{Ising_mapping, D-Wave_application6, D-Wave_application16}. Accordingly, one-hot encoding is one of the most commonly used approaches.
This subsection briefly explains an Ising formulation of integer optimization problems with one-hot encoding.

\begin{figure}
\centering
\includegraphics[width=0.5\columnwidth]{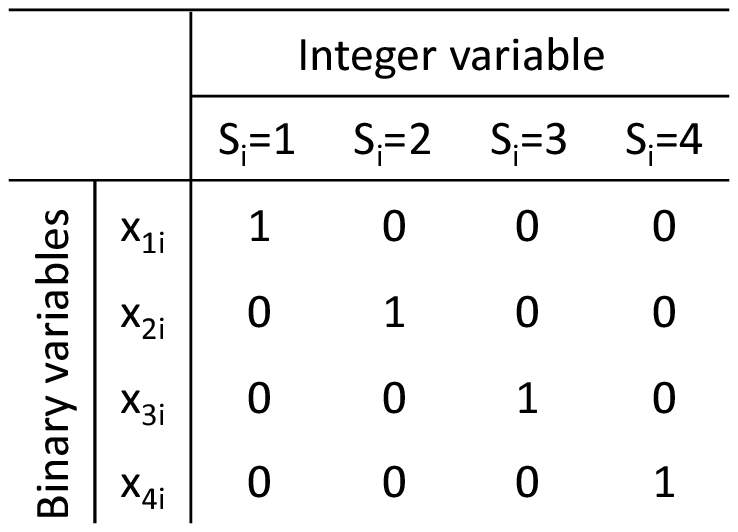}
\caption{Binary representation of $S_{i}$ with one-hot encoding.}
\label{fig:formulation_conv_binary2integer}
\end{figure}

A general-form integer optimization problem is given as follows:
\begin{equation}
\min_{\bm{S}} \left[ - \sum_{(ij) \in E} f_{ij}(S_{i}, S_{j}) - \sum_{i=1}^{N} g_{i}(S_{i}) \right],
\end{equation}
where $\bm{S} = \{ S_{1}, S_{2}, ..., S_{N} \}$, $S_{i} \in (1, 2, ..., Q)$ is an integer variable; $N$ is the number of integer variables; $f_{ij}(S_{i}, S_{j})$ is the interaction between $S_{i}$ and $S_{j}$; $(i,j) \in E$ is a pair of indices of integer variables with the interaction; and $g_{i}(S_{i})$ is an external magnetic field applied to $S_{i}$.
One-hot encoding assigns $Q$ binary variables $\{ x_{qi} \in (0,1) | q =1, 2, ..., Q \}$ to $S_{i}$.
The following constraint
\begin{equation}
\sum_{q = 1}^{Q} x_{qi} = 1, \ \ \forall i,
\end{equation}
is imposed, and $S_{i}$ is restored as follows:
\begin{equation}
S_{i} = \sum_{q=1}^{Q} q x_{qi}.
\end{equation}
Fig. \ref{fig:formulation_conv_binary2integer} for $Q=4$ presents the binary representation of an integer variable with one-hot encoding.
The equivalent optimization problem defined by binary variables is given as
\begin{eqnarray}
\min_{\bm{x}} \left[ - \sum_{(ij) \in E} \sum_{q=1}^{Q} \sum_{q'=1}^{Q} f_{ij}(q,q') x_{qi} x_{q'j} - \sum_{i=1}^{N} \sum_{q=1}^{Q} g_{i}(q) x_{qi} \right]&&  \nonumber  \\
\mathrm{s.t.} \ \sum_{q=1}^{Q} x_{qi} = 1. \ \ \ \ \ &&  \label{eq:formulation_Conv_binary}
\end{eqnarray}
Using the following transformation:
\begin{equation}
x_{qi} = \frac{ 1 - \sigma^{z}_{qi} }{2},  \label{eq:formulation_binary2sigma}
\end{equation}
the abovementioned optimization problem [Eq. (\ref{eq:formulation_Conv_binary})] is expressed as follows with respect to the Ising spin $\sigma_{qi} \in (+1, -1)$:
\begin{eqnarray}
\min_{\bm{\sigma}^{z}} \left[ - \sum_{(ij) \in E} \sum_{q=1}^{Q} \sum_{q'=1}^{Q} J_{ij}(q,q') \sigma^{z}_{qi} \sigma^{z}_{q'j} - \sum_{i=1}^{N} \sum_{q=1}^{Q} h_{i}(q) \sigma^{z}_{qi} \right]&& \nonumber  \\
\mathrm{s.t.} \ \sum_{q=1}^{Q} \sigma^{z}_{qi} = Q-2, \ \ \ \ \ &&
\end{eqnarray}
where
\begin{equation}
J_{ij}(q,q') = \frac{1}{4} f_{ij}(q,q'),
\end{equation}
and
\begin{equation}
h_{i}(q) = - \frac{1}{4} \sum_{j \in \partial i} \sum_{q'=1}^{Q} f_{ij}(q,q') - \frac{1}{2} g_{i}(q).
\end{equation}
Here, $\sigma_{qi} = -1$ represents $S_{i} = q$, and $j \in \partial i$ represents the indices of the integer variables that interact with $S_{i}$.
An unconstrained cost function is required to solve the above optimization problem using Ising-type computers.
We obtain the problem Hamiltonian as follows by introducing a penalty term:
\begin{eqnarray}
\mathcal{H}_{0} = &-& \sum_{(ij) \in E} \sum_{q=1}^{Q} \sum_{q'=1}^{Q} J_{ij}(q,q') \sigma^{z}_{qi} \sigma^{z}_{q'j} - \sum_{i=1}^{N} \sum_{q=1}^{Q} h_{i}(q) \sigma^{z}_{qi}  \nonumber  \\
&+& \frac{\lambda}{2} \sum_{i=1}^{N} \left( \sum_{q=1}^{Q} \sigma^{z}_{qi} - (Q-2) \right)^{2},
\end{eqnarray}
where the third term represents the penalty term, and $\lambda$ controls the strength of the penalty term.
By using one-hot encoding, we can express integer optimization problems as a ground state search of an Ising model for arbitrary $f_{ij}(S_{i},S_{j})$.

\subsection{Proposed formulation}
We modify one-hot encoding to embed information of a candidate solution into the problem Hamiltonian.

\begin{figure}
\centering
\includegraphics[width=0.5\columnwidth]{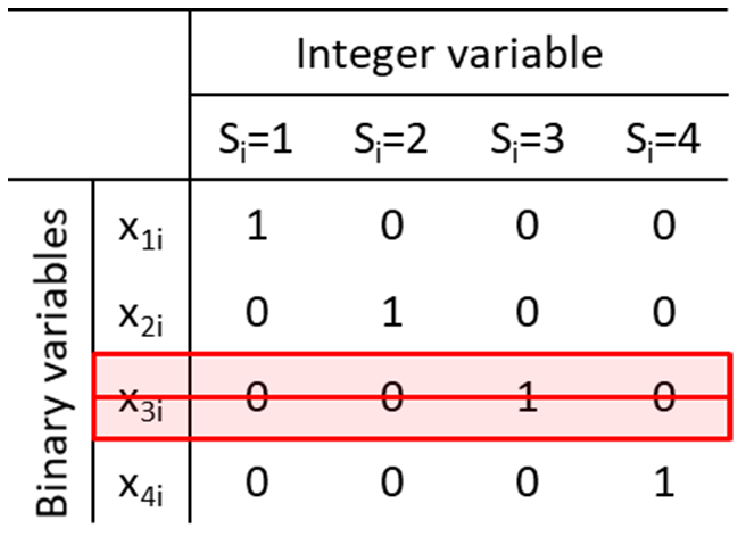}
\caption{Binary representation of $S_{i}$ with the proposed formulation.
When a candidate solution is $S_{i} = 3$, $x_{3i}$ is removed from conventional one-hot encoding.}
\label{fig:formulation_prop_binary2integer}
\end{figure}

Given that we have a candidate solution: $S_{i} = r_{i}, \forall i$.
Our formulation removes $x_{r_{i}, i}$ from the conventional one-hot encoding.
Fig. \ref{fig:formulation_prop_binary2integer} depicts the binary representation of $S_{i}$ with the proposed formulation for $Q=4$ and $r_{i} = 3$.
While the candidate solution $S_{i} = r_{i}$ is represented as a \textit{zero-hot} configuration, $S_{i} \neq r_{i}$ is represented as a one-hot configuration.
To add $\{ x_{qi} = 0 | q \neq r_{i} \}$ to feasible solutions, the constraint imposed in the conventional one-hot encoding is modified as follows:
\begin{equation}
\sum_{q \neq r_{i}} x_{qi} = 0 \ \mathrm{or} \  1.  \label{eq:formulation_prop_constraint}
\end{equation}
By substituting
\begin{equation}
x_{r_{i}, i} = 1 - \sum_{q \neq r_{i}} x_{qi},
\end{equation}
into Eq. (\ref{eq:formulation_Conv_binary}), in addition to changing the constraint to Eq. (\ref{eq:formulation_prop_constraint}), we obtain the optimization problem expressed with respect to the binary variable $x_{qi} \in (0,1)$ as follows:
\begin{eqnarray}
\min_{\bm{x}} \left[ - \sum_{(ij) \in E} \sum_{q \neq r_{i}} \sum_{q' \neq r_{j}} Q_{ij}(q,q') x_{qi} x_{q'j} - \sum_{i} \sum_{q \neq r_{i}} R_{i}(q) x_{qi} \right]&&  \nonumber  \\
\mathrm{s.t.} \  \sum_{q \neq r_{i}} x_{qi} = 0 \ \mathrm{or} \ 1, \ \ \ \ \ \ \ &&
\end{eqnarray}
where
\begin{eqnarray}
Q_{ij}(q,q') &=& f_{ij}(q,q') - f_{ij}(q,r_{i})  \nonumber  \\
&-& f_{ij}(r_{i},q') + f_{ij}(r_{i},r_{j}),  \label{eq:formulation_prop_Qij}
\end{eqnarray}
and
\begin{eqnarray}
R_{i}(q) &=& \sum_{j \in \partial i} f_{ij}(q,r_{j}) - \sum_{j \in \partial i} f_{ij}(r_{i},r_{j})  \nonumber  \\
&+& g_{i}(q) - g_{i}(r_{i}).  \label{eq:formulation_prop_Ri}
\end{eqnarray}
Eq. (\ref{eq:formulation_binary2sigma}) then yields
\begin{eqnarray}
\min_{\bm{\sigma}^{z}} \left[ - \sum_{(ij) \in E} \sum_{q \neq r_{i}} \sum_{q' \neq r_{j}} J_{ij}(q,q') \sigma^{z}_{qi} \sigma^{z}_{q'j} - \sum_{i} \sum_{q \neq r_{i}} h_{i}(q) \sigma^{z}_{qi} \right]&&  \nonumber  \\
\mathrm{s.t.} \  \sum_{q \neq r_{i}} \sigma^{z}_{qi} = Q-1 \ \mathrm{or} \  Q-3, \ \ \ \ \ \ \ &&
\end{eqnarray}
where
\begin{equation}
J_{ij}(q,q') = \frac{1}{4} Q_{ij}(q,q'),  \label{eq:formulation_prop_Jij}
\end{equation}
and
\begin{equation}
h_{i}(q) = - \frac{1}{4} \sum_{j \in \partial i} \sum_{q' \neq r_{j}} Q_{ij}(q,q') - \frac{1}{2} R_{i}(q).  \label{eq:formulation_prop_hi}
\end{equation}
Finally, the unconstrained cost function is obtained by introducing the following penalty term:
\begin{eqnarray}
\mathcal{H}_{0} = &-& \sum_{(ij) \in E} \sum_{q \neq r_{i}} \sum_{q' \neq r_{j}} J_{ij}(q,q') \sigma^{z}_{qi} \sigma^{z}_{q'j} - \sum_{i} \sum_{q \neq r_{i}} h_{i}(q) \sigma^{z}_{qi}  \nonumber  \\
&+& \frac{\lambda}{2} \sum_{i} \left( \sum_{q \neq r_{i}} \sigma^{z}_{qi} - (Q-2) \right)^{2}.  \label{eq:formulation_prop_Ising_Ham}
\end{eqnarray}
The penalty term is minimized when $\sum_{q \neq r_{i}} \sigma^{z}_{qi} = Q-1 \ \mathrm{or} \  Q-3$.

\subsection{QA of the penalty term of the proposed formulation}
The proposed formulation embeds information about a candidate solution into the problem Hamiltonian by exploiting the biased sampling of the degenerated ground states in TMF-QA.
By analyzing QA of the penalty term, we confirm in this section that the phase space around the candidate solution is mainly searched with the proposed formulation.

\begin{figure}
\centering
\includegraphics[width=0.5\columnwidth]{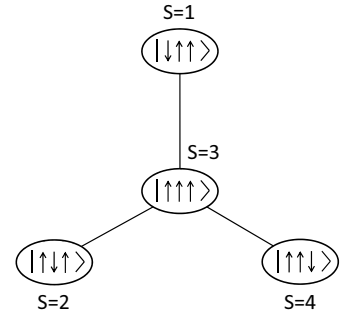}
\caption{Ground states of the penalty term for $Q=4$ and $r=3$.
The upward and downward arrows indicate $\sigma^{z}_{q} = +1$ and $\sigma^{z}_{q} =-1$, respectively.
For example, $\ket{\downarrow \uparrow  \uparrow}$ means $\sigma^{z}_{1} = -1, \sigma^{z}_{2} = \sigma^{z}_{4} = +1$.
Note that $\sigma^{z}_{3}$ is removed because $r = 3$.}
\label{fig:formulation_prop_pen_GS}
\end{figure}

The Hamiltonian of TMF-QA is given as follows:
\begin{eqnarray}
\hat{\mathcal{H}}_{\mathrm{QA}} &=& \hat{\mathcal{H}}_{\mathrm{pen}} - \Gamma \sum_{q \neq r} \hat{\sigma}^{x}_{q},  \\
\hat{\mathcal{H}}_{\mathrm{pen}} &=& \frac{\lambda}{2} \left( \sum_{q \neq r} \hat{\sigma}^{z}_{q} - (Q-2) \right)^{2},  \label{eq:formulation_prob_Ham_pen}
\end{eqnarray}
where $\hat{\mathcal{H}}_{\mathrm{pen}}$ is the penalty term of the proposed formulation, and $\sigma^{z}_{q}, \sigma^{x}_{q}$ are the Pauli $z, x$ operators.
Fig. \ref{fig:formulation_prop_pen_GS} depicts the ground states of the penalty term for $Q=4$ and $r=3$, where the Hamming distance between the ground states connected by an edge is one.
Hereafter, we refer to the ground state satisfying $\sum_{q \neq r} \sigma^{z}_{qi} = Q-3$ as the zero-hot ground state and that satisfying $\sum_{q \neq r} \sigma^{z}_{qi} = Q-1$ as the one-hot ground state.

The biased sampling of the degenerated ground states contributes to the information embedding about the candidate solution into the problem Hamiltonian.
Free spins play an important role in causing the biased sampling in TMF-QA \cite{unfair_sampling1, unfair_sampling2}.
Here, a free spin is referred to as a spin that does not change the energy by flipping it in the ground states.
The number of free spins is different between the zero- and one-hot ground states: $Q-1$ for the zero-hot ground state and one for each one-hot ground state.
When $\Gamma$ is small, the ground state of $\hat{\mathcal{H}}_{\mathrm{QA}}$ can be approximately expressed as a superposition of the ground states of $\hat{\mathcal{H}}_{\mathrm{pen}}$.
To lower the energy term of the transverse magnetic field, the zero- and one-hot ground states must be superposed to obtain the eigen vector of $\hat{\sigma}^{x}_{q}$.
For example, when $Q=4$, the following three states
\begin{eqnarray}
\ket{ \rightarrow \uparrow \uparrow } &=& \frac{1}{\sqrt{2}} \left( \ket{ \downarrow \uparrow \uparrow } + \ket{ \uparrow \uparrow \uparrow } \right),  \\
\ket{ \uparrow \rightarrow \uparrow } &=& \frac{1}{\sqrt{2}} \left( \ket{ \uparrow \downarrow \uparrow } + \ket{ \uparrow \uparrow \uparrow } \right),  \\
\ket{ \uparrow \uparrow \rightarrow } &=& \frac{1}{\sqrt{2}} \left( \ket{ \uparrow \uparrow \downarrow } + \ket{ \uparrow \uparrow \uparrow } \right),
\end{eqnarray}
contain the eigen vector of $\hat{\sigma}^{x}_{q}$.
Here, the right arrow ``$\rightarrow$'' indicates that a spin is aligned along the transverse magnetic field.
The zero-hot ground state is commonly used to obtain the eigen vector of $\hat{\sigma}^{x}_{q}$.
Consequently, the zero-hot ground state is sampled with a higher probability than each one-hot ground state.
Using the degenerate perturbation theory, the probability of obtaining the zero-hot ground state can easily be verified as 50\% at $\Gamma = 0$ \cite{unfair_sampling4}.

\begin{figure}
\centering
\includegraphics[width=0.9\columnwidth]{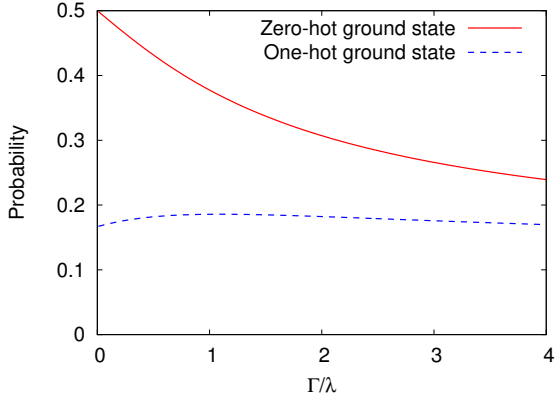}
\caption{Probability of obtaining the zero- and one-hot ground states during QA.}
\label{fig:formulation_prop_pen_QA_probability}
\end{figure}

Fig. \ref{fig:formulation_prop_pen_QA_probability} shows the dependence of $P(S=r)$ and $P(S = r')$ for $r' \neq r$ on $\Gamma$ during QA.
Here, $P(S=r)$ and $P(S = r')$ represent the probability of obtaining $S=r$ and $S = r'$, respectively. We set $Q=4$.
The probability of obtaining the candidate solution $P(S=r)$ is always greater than $P(S = r')$. In other words, the phase space around the candidate solution is mainly searched during QA with the proposed formulation.
As explained in the previous paragraph, $P(S=r) = 0.5$ at $\Gamma = 0$.
Here, note that the proposed formulation cannot be applied to the optimization by SA for embedding the candidate solution 
because the degenerated ground states are fairly sampled in SA.

\section{QA of the FC-FM Potts model with the proposed formulation}
We apply the proposed formulation to QA of the FC-FM Potts model.
QA with conventional one-hot encoding was analytically investigated in Ref. \cite{half-hot_optimization}, which confirmed the occurrence of a first-order phase transition.

\subsection{Ising formulation with the proposed method}
The Hamiltonian of the FC-FM Potts model is given as follows:
\begin{equation}
\mathcal{H}_{\mathrm{potts}}(\bm{S}) = - \sum_{i<j} f_{ij}(S_{i},S_{j}) - \sum_{i} g_{i}(S_{i}),
\end{equation}
\begin{equation}
f_{ij}(S_{i},S_{j}) = \frac{4J}{N} \delta (S_{i},S_{j}),  \label{eq:FM_Ham_fij_def}
\end{equation}
\begin{equation}
g_{i}(S_{i}) = 0,  \label{eq:FM_Ham_gi_def}
\end{equation}
where $J>0$ represents the strength of the ferromagnetic interaction, and $\delta$ denotes the Kronecker delta function.
From Eqs. (\ref{eq:formulation_prop_Qij}), (\ref{eq:formulation_prop_Ri}), (\ref{eq:formulation_prop_Jij}), (\ref{eq:formulation_prop_hi}), (\ref{eq:formulation_prop_Ising_Ham}), (\ref{eq:FM_Ham_fij_def}), and (\ref{eq:FM_Ham_gi_def}), we obtain the Ising formulation of the Hamiltonian of the FC-FM Potts model as follows:
\begin{eqnarray}
\mathcal{H}_{0} = &-& \sum_{i<j} \sum_{q \neq r_{i}} \sum_{q' \neq r_{j}} J_{ij}(q,q') \sigma^{z}_{qi} \sigma^{z}_{q'j} - \sum_{i} \sum_{q \neq r_{i}} h_{i}(q) \sigma^{z}_{qi}  \nonumber  \\
&+& \frac{\lambda}{2} \sum_{i} \left( \sum_{q \neq r_{i}} \sigma^{z}_{qi} - (Q-2) \right)^{2}.  \label{eq:formulation_prop_Ising_Ham2}
\end{eqnarray}
where
\begin{equation}
J_{ij}(q,q') = \frac{J}{N} \left[ \delta (q,q') - \delta (q,r_{j}) - \delta (r_{i},q') + \delta (r_{i},r_{j}) \right],
\end{equation}
and
\begin{equation}
h_{i}(q) = \frac{Q-2}{N} J \sum_{j \neq i} \left[ \delta (q,r_{i}) - \delta (r_{i},r_{j}) \right].
\end{equation}

\begin{figure}
\centering
\includegraphics[width=0.95\columnwidth]{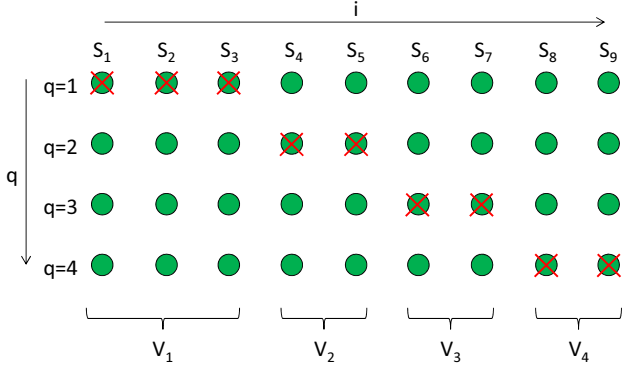}
\caption{Example of the proposed formulation.
The vertices represent the spin variables $\{ \sigma_{qi} \}$. Four vertically arranged spins are assigned to the same integer variable in the conventional formulation.
In the proposed formulation, $\sigma_{11}$, $\sigma_{12}$, and $\sigma_{13}$ are removed from the conventional formulation because $S_{1}$, $S_{2}$, and $S_{3}$ belong to $V_{1}$ and $r_{1} = r_{2} = r_{3} = 1$.
In the same manner, $\{ \sigma_{\xi i} | i \in V_{\xi} \}$ is removed for the other groups ($\xi > 1$) because the candidate solution is $S_{i} = \xi$ for $i \in V_{\xi}$.}
\label{fig:FM_prop_formulation}
\end{figure}

Next, we divide the $N$ integer variables $\{ S_{i} \}_{i = 1, 2, ..., N}$ into $Q$ groups $\{ V_{\xi} \}_{\xi = 1, 2, ..., Q}$, where a candidate solution is given by $\{ r_{i} = \xi | i \in V_{\xi} \}$ for each group.
Fig. \ref{fig:FM_prop_formulation} depicts a simple example for $Q=4$ and $N=9$.
The vertices represent the spin variables $\{ \sigma_{qi} \}$. Four spins vertically arranged are assigned to the same integer variable.
In this example, $\sigma_{11}, \sigma_{12}$, and $\sigma_{13}$ are removed because $S_{1}$, $S_{2}$, and $S_{3}$ belong to $V_{1}$, and $r_{1} = r_{2} = r_{3} = 1$.
In the same manner, $\{ \sigma_{\xi i} | i \in V_{\xi} \}$ is removed for other groups ($\xi > 1$) because the candidate solution is $S_{i} = \xi$ for $i \in V_{\xi}$.
By expanding the summation over $i$ as
\begin{equation}
\sum_{i} = \sum_{\xi} \sum_{i \in V_{\xi}},
\end{equation}
we can rewrite the Hamiltonian [Eq. (\ref{eq:formulation_prop_Ising_Ham2})] as
\begin{widetext}
\begin{eqnarray}
\mathcal{H}_{0}(\bm{\sigma}^{z}) = &&- \frac{N}{2} \sum_{\xi = 1}^{Q} \sum_{\eta =1}^{Q} \rho_{\xi} \rho_{\eta} \sum_{q \neq \xi} \sum_{q' \neq \eta} \tilde{J}_{\xi \eta}(q,q') \left( \frac{1}{N_{\xi}} \sum_{i \in V_{\xi}} \sigma^{z}_{qi} \right) \left( \frac{1}{N_{\eta}} \sum_{j \in V_{\eta}} \sigma^{z}_{q'j} \right)  \nonumber  \\
&&- (Q-2) \sum_{\xi} \sum_{i \in V_{\xi}} \sum_{q \neq \xi} \tilde{h}_{\xi}(q) \sigma^{z}_{qi} + \frac{\lambda}{2} \sum_{\xi} \sum_{i \in V_{\xi}} \left( \sum_{q \neq \xi} \sigma^{z}_{qi} - (Q-2) \right)^{2},  \label{eq:FM_prop_problem_Ham}
\end{eqnarray}
\end{widetext}
where $N_{\xi}$ is the number of integer variables belonging to $V_{\xi}$, $\rho_{\xi} = N_{\xi} / N$, and $\tilde{J}_{\xi \eta}(q,q'), \tilde{h}_{\xi}(q)$ are given as
\begin{equation}
\tilde{J}_{\xi \eta}(q,q') = J \left[ \delta (q,q') - \delta (q, \eta) - \delta (\xi, q') + \delta (\xi, \eta) \right],
\end{equation}
\begin{equation}
\tilde{h}_{\xi}(q) = J(\rho_{q} - \rho_{\xi}).
\end{equation}
Hereafter, we set
\begin{equation}
\rho_{1} > \rho_{\xi \geq 2},
\end{equation}
In this case, $\rho_{1}$ represents the overlap between the candidate solution $\{ S_{i} = \xi | i \in V_{\xi}, \xi = 1, 2, ..., Q \}$ and the ground state $S_{i} = 1$ of the FC-FM Potts model.
We will analytically investigate the order of the phase transition during TMF-QA for various $\rho_{1}$ values in the subsequent sections.

\subsection{Free energy and saddle point equations}
\begin{figure}
\centering
\includegraphics[width=0.95\columnwidth]{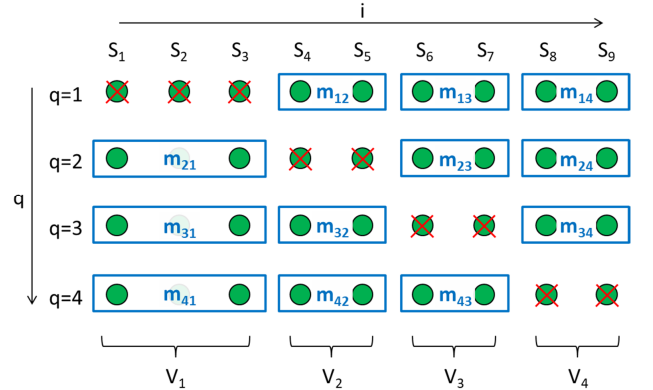}
\caption{Ferromagnetic order parameters introduced in the proposed formulation.}
\label{fig:FM_prop_order_parameters}
\end{figure}

The Hamiltonian of TMF-QA is given as follows:
\begin{equation}
\hat{\mathcal{H}}_{\mathrm{QA}} = \hat{\mathcal{H}}_{0}(\hat{\bm{\sigma}}^{z}) + \hat{\mathcal{H}}_{q}(\hat{\bm{\sigma}}^{x}),
\end{equation}
\begin{equation}
\hat{\mathcal{H}}_{q}(\hat{\bm{\sigma}}^{x}) = - \Gamma \sum_{\xi} \sum_{i \in V_{\xi}} \sum_{q \neq \xi} \hat{\sigma}^{x}_{qi},  \label{eq:FM_prop_driver_Ham}
\end{equation}
where $\mathcal{H}_{0}(\bm{\sigma}^{z})$ is given by Eq. (\ref{eq:FM_prop_problem_Ham}).
By applying the Suzuki--Trotter formula \cite{ST_formula} and assuming the static approximation \cite{static_approx_exact} that neglects the imaginary-time dependence of the order parameters,
we obtain free energy as follows in the limit of $N \to \infty$:
\begin{widetext}
\begin{equation}
f(\bm{m}, \tilde{\bm{m}}) = - \frac{1}{2} \sum_{\xi, \eta} \rho_{\xi} \rho_{\eta} \sum_{q \neq \xi} \sum_{q' \neq \eta} \tilde{J}_{\xi \eta}(q,q') m_{q \xi} m_{q' \eta} + J \sum_{\xi} \sum_{q \neq \xi} \rho_{\xi} \tilde{m}_{q \xi} m_{q \xi} - \frac{1}{\beta} \sum_{\xi} \rho_{\xi} \log \mathrm{Tr} e^{- \beta \hat{\mathcal{H}}^{(\mathrm{eff})}_{\xi}},  \label{eq:FM_prop_free_energy}
\end{equation}
\end{widetext}
where $\beta$ is the inverse temperature; $m_{q \xi}$ is the ferromagnetic order parameter for $\{ \sigma^{z}_{qi} | i \in V_{\xi} \}$(Fig. \ref{fig:FM_prop_order_parameters}); $\tilde{m}_{q \xi}$ is a conjugate variable of $m_{q \xi}$; and the effective Hamiltonian $\hat{\mathcal{H}}^{(\mathrm{eff})}_{\xi}$ is given as
\begin{equation}
\hat{\mathcal{H}}^{(\mathrm{eff})}_{\xi} = \frac{\lambda}{2} \left( \sum_{q \neq \xi} \hat{\sigma}^{z}_{q} \right)^{2} - \sum_{q \neq \xi} \tilde{h}^{(\mathrm{eff})}_{\xi}(q) \hat{\sigma}^{z}_{q} - \Gamma \sum_{q \neq \xi} \hat{\sigma}^{x}_{q},  \label{eq:FM_prop_Ham_eff_def}
\end{equation}
and
\begin{equation}
\frac{\tilde{h}^{(\mathrm{eff})}_{\xi}(q)}{J} = \tilde{m}_{q \xi} + (Q-2) \left( \rho_{q} - \rho_{\xi} + \frac{\lambda}{J} \right).  \label{eq:FM_prop_h_eff_def}
\end{equation}
Appendix A presents a detailed derivation of the free energy.
The order parameters are determined as the free energy minimizer. Saddle point equations are given below:
\begin{equation}
\tilde{m}_{q \xi} = \frac{1}{J} \sum_{\eta} \rho_{\eta} \sum_{q' \neq \eta} \tilde{J}_{\xi \eta}(q,q') m_{q' \eta},  \label{eq:FM_prop_saddle_point_eq1}
\end{equation}
\begin{equation}
m_{q \xi} = \frac{ \displaystyle{ \mathrm{Tr} \hat{\sigma}^{z}_{q} e^{- \beta \hat{\mathcal{H}}^{(\mathrm{eff})}_{\xi}(\hat{\bm{\sigma}}^{z,x}, \tilde{\bm{m}}_{\xi}) } } }{ \displaystyle{ \mathrm{Tr} e^{- \beta \hat{\mathcal{H}}^{(\mathrm{eff})}_{\xi}(\hat{\bm{\sigma}}^{z,x}, \tilde{\bm{m}}_{\xi}) } } } \equiv \left\langle \hat{\sigma}^{z}_{q} \right\rangle_{\hat{\mathcal{H}}^{(\mathrm{eff})}_{\xi}}.  \label{eq:FM_prop_saddle_point_eq2}
\end{equation}

\subsection{With the candidate solution identical to one of the ground states: $\rho_{1} = 1, \rho_{\xi \geq 2} = 0$}
In this subsection, we investigate the phase transition order in the case of one of the ground states $S_{i} = 1$ given as a candidate solution.
We particularly confirm that the first-order phase transition remains in SA while being successfully avoided in QA.
This result implies that for the SA optimization, the proposed formulation cannot be used to embed information about the candidate solution into the problem Hamiltonian.

\begin{figure}
\centering
\includegraphics[width=0.95\columnwidth]{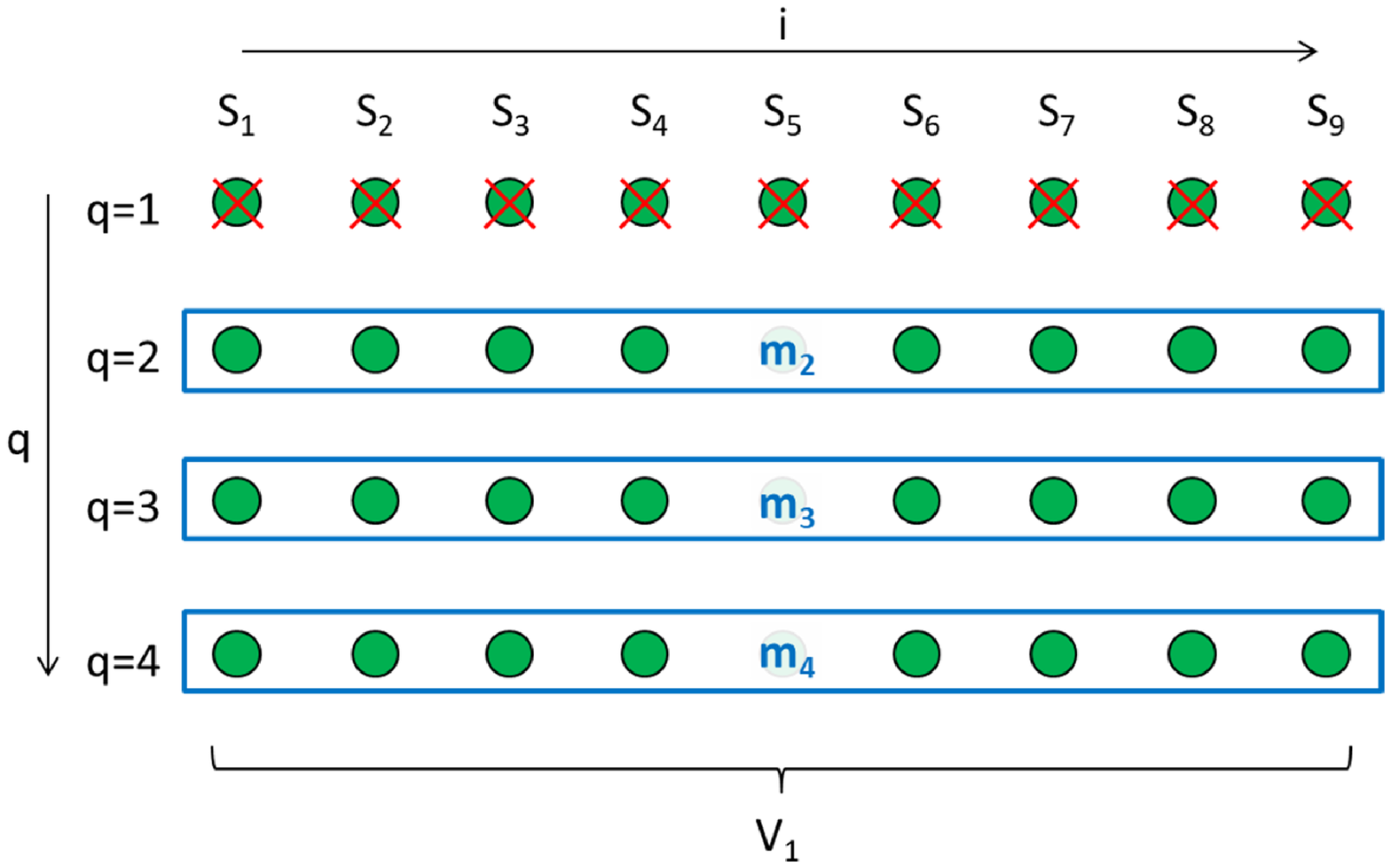}
\caption{Ferromagnetic order parameters introduced in the proposed formulation.}
\label{fig:FM_prop_GS_order_parameters}
\end{figure}

The free energy is given as follows:
\begin{eqnarray}
f(\bm{m}, \tilde{\bm{m}}) = &-& \frac{J}{2} \sum_{q \neq 1} \sum_{q' \neq 1} \left[ 1 + \delta (q,q') \right] m_{q} m_{q'}  \nonumber  \\
&+& J \sum_{q \neq 1} \tilde{m}_{q} m_{q} - \frac{1}{\beta} \log \mathrm{Tr} e^{- \beta \hat{\mathcal{H}}^{(\mathrm{eff})}},
\end{eqnarray}
where
\begin{equation}
\hat{\mathcal{H}}^{(\mathrm{eff})} = \frac{\lambda}{2} \left( \sum_{q \neq 1} \hat{\sigma}^{z}_{q} \right)^{2} - \sum_{q \neq 1} \tilde{h}^{(\mathrm{eff})}(q) \hat{\sigma}^{z}_{q} - \Gamma \sum_{q \neq 1} \hat{\sigma}^{x}_{q},
\end{equation}
and 
\begin{equation}
\frac{\tilde{h}^{(\mathrm{eff})}(q)}{J} = \tilde{m}_{q} + (Q-2) \left( \frac{\lambda}{J} - 1 \right).
\end{equation}
$\rho_{\xi \geq 2} = 0$; hence, the free energy is defined by the $Q-1$ ferromagnetic order parameters $\{ m_{q} | q \neq 1 \}$ and $Q-1$ conjugate variables $\{ \tilde{m}_{q} | q \neq 1 \}$(Fig. \ref{fig:FM_prop_GS_order_parameters}).
The saddle point equations are given as follows:
\begin{equation}
\tilde{m}_{q} = \sum_{q' \neq 1} \left[ 1 + \delta (q,q') \right] m_{q'},
\end{equation}
\begin{equation}
m_{q} = \left\langle \hat{\sigma}^{z}_{q} \right\rangle_{\hat{\mathcal{H}}^{(\mathrm{eff})}}.
\end{equation}

\begin{figure}
\centering
\includegraphics[width=0.8\columnwidth]{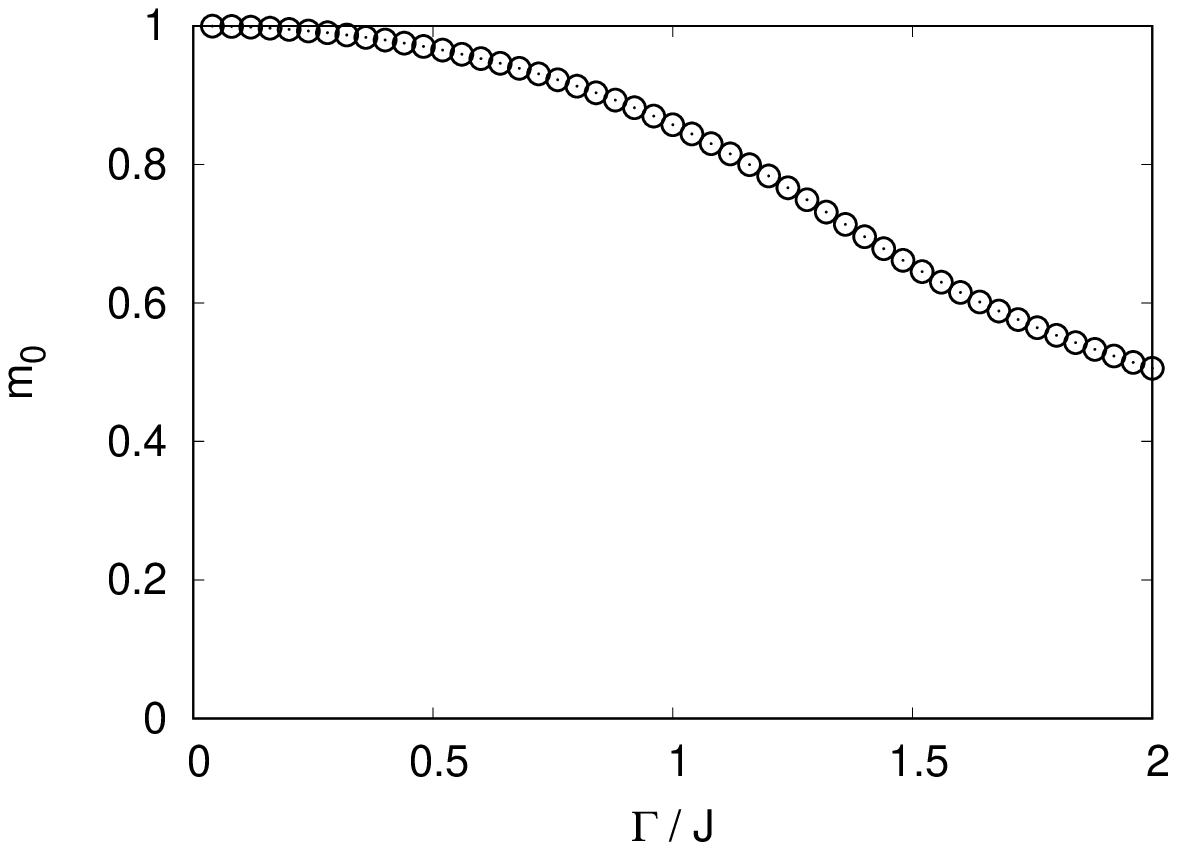}
\caption{Dependence of $m_{0}$ on $\Gamma$ at $T=0$. We set $\rho_{1} = 1, \lambda = 1.5$.}
\label{fig:FM_rho1=1_QA_m0}
\end{figure}
\begin{figure}
\centering
\includegraphics[width=0.8\columnwidth]{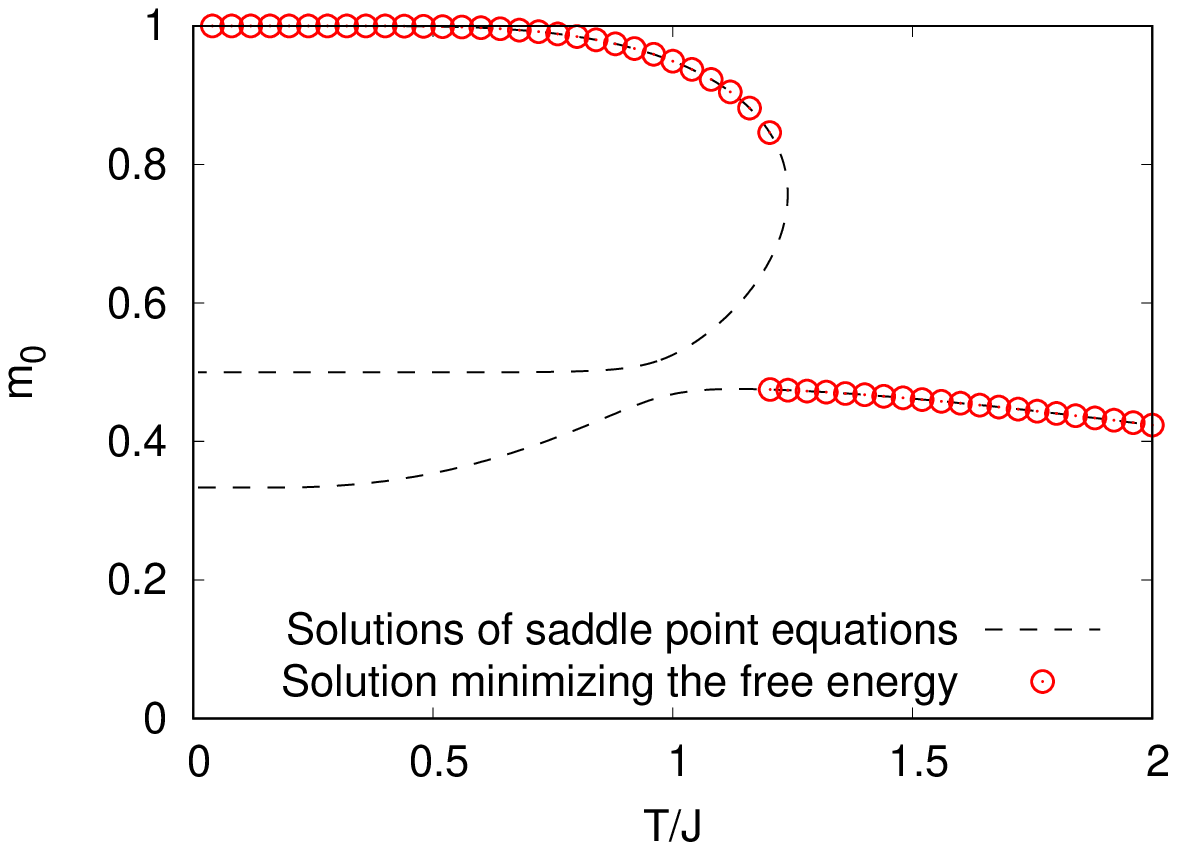}
\caption{Dependence of $m_{0}$ on $T$ at $\Gamma=0$. We set $\rho_{1} = 1, \lambda = 1.5$.}
\label{fig:FM_rho1=1_SA_m0}
\end{figure}
\begin{figure}[h]
\centering
\includegraphics[width=0.8\columnwidth]{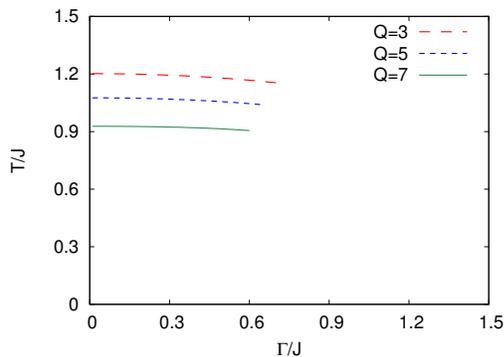}
\caption{Condition where the first-order phase transition occurs on the $\Gamma - T$ plane for $Q=3, 5$, and $7$. We set $\rho_{1} = 1, \lambda = 1.5$.}
\label{fig:FM_rho1=1_SQA_phase_diagram}
\end{figure}
The ferromagnetic order parameters and the conjugate variables at the global minimum of the free energy do not depend on $q$:
\begin{eqnarray}
m_{0} &=& m_{q},  \\
\tilde{m}_{0} &=& \tilde{m}_{q}.
\end{eqnarray}
Fig. \ref{fig:FM_rho1=1_QA_m0} depicts the dependence of $m_{0}$ on $\Gamma$ at $T=0$ for $Q=4$, where $T = 1 / \beta$ denotes temperature.
The ferromagnetic order parameter continuously changes as $\Gamma$ decreases, indicating that the first-order phase transition during QA is avoided.
Fig. \ref{fig:FM_rho1=1_SA_m0} illustrates the dependence of $m_{0}$ on $T$ at $\Gamma = 0$ for $Q=4$.
The dashed lines represent the solutions of the saddle point equations. The points are solutions that minimize the free energy.
A metastable state appears as $T$ decreases. A discontinuous jump of $m_{0}$ also occurs.
In other words, the first-order phase transition remains during SA.
Fig. \ref{fig:FM_rho1=1_SQA_phase_diagram} depicts a phase diagram of simulated QA on the $\Gamma - T$ plane for $Q=3, 5,$ and $7$.
Here, we set $\rho_{1} = 1$ and $\lambda = 1.5$.
The lines in Fig. \ref{fig:FM_rho1=1_SQA_phase_diagram} represent the condition where the first-order phase transition occurs for each $Q$ value.
As the effect of the biased sampling is made stronger by the $\Gamma$ increase, the discontinuous jump of $m_{0}$ becomes smaller, and the first-order phase transition finally disappears.
Considering the abovementioned results, we conclude that the proposed formulation is applicable to the TMF-QA optimization for embedding information about the candidate solution into the problem Hamiltonian.

\begin{figure}
\centering
\includegraphics[width=0.5\columnwidth]{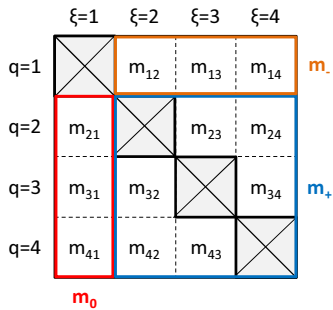}
\caption{Symmetry of the ferromagnetic order parameters when $\rho_{\xi \geq 2} = \rho$.
We set $Q=4$.}
\label{fig:FM_mag_symmetry}
\end{figure}

\begin{figure*}
\centering
\begin{subfigure}[b]{0.8\columnwidth}
\centering
\includegraphics[width=\columnwidth]{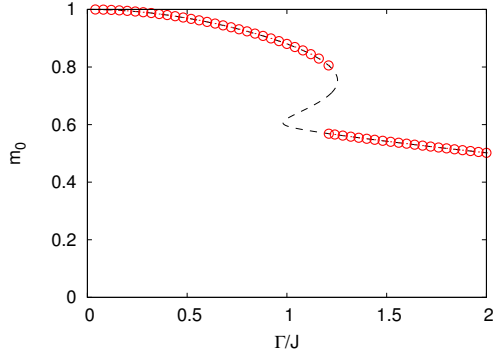}
\caption{Dependence of $m_{0}$ on $\Gamma$ for $\rho_{1} = 0.26$.}
\end{subfigure}
\hspace{50pt}
\begin{subfigure}[b]{0.8\columnwidth}
\centering
\includegraphics[width=\columnwidth]{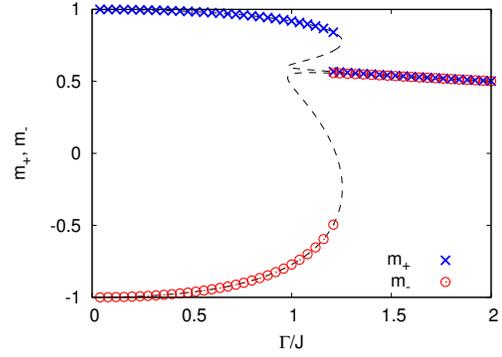}
\caption{Dependence of $m_{\pm}$ on $\Gamma$ for $\rho_{1} = 0.26$.}
\end{subfigure}
\caption{Dependence of $m_{0}$ and $m_{\pm}$ on $\Gamma$ for $\rho_{1} = 0.26$.
The dashed lines represent the solutions of the saddle point equations (\ref{eq:FM_prop_saddle_point_eq1}) and (\ref{eq:FM_prop_saddle_point_eq2}).
The points are one of the solutions minimizing the free energy.
We set $Q=4$ and $\lambda = 1.5$.}
\label{fig:FM_rho1=0.26_mag}
\end{figure*}

\begin{figure*}
\centering
\begin{subfigure}[b]{0.8\columnwidth}
\centering
\includegraphics[width=\columnwidth]{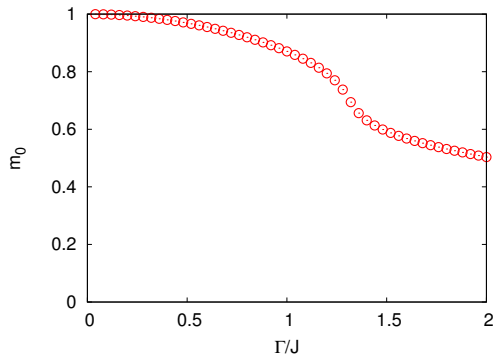}
\caption{Dependence of $m_{0}$ on $\Gamma$ for $\rho_{1} = 0.60$.}
\end{subfigure}
\hspace{50pt}
\begin{subfigure}[b]{0.8\columnwidth}
\centering
\includegraphics[width=\columnwidth]{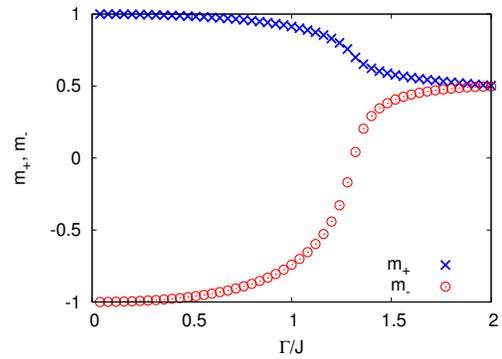}
\caption{Dependence of $m_{\pm}$ on $\Gamma$ for $\rho_{1} = 0.60$.}
\end{subfigure}
\caption{Dependence of $m_{0}$ and $m_{\pm}$ on $\Gamma$ for $\rho_{1} = 0.60$.
We set $Q=4$ and $\lambda = 1.5$.}
\label{fig:FM_rho1=0.6_mag}
\end{figure*}
\subsection{With the candidate solution different from the ground states: $1 > \rho_{1} > \rho_{\xi \geq 2} = \rho$}
We next investigate the phase transition order during TMF-QA ($T=0$) for various $\rho_{1}$. We confirm that the first-order phase transition is successfully avoided when $\rho_{1}$ exceeds a threshold.
In this subsection, we set
\begin{equation}
\rho_{\xi \geq 2} = \frac{1 - \rho_{1}}{Q-1} \equiv \rho.
\end{equation}

The ferromagnetic order parameters satisfy the following equations at the global minimum of the free energy:
\begin{eqnarray}
m_{0} &=& m_{q1}, q \geq 2,  \\
m_{-} &=& m_{1 \xi}, \xi \geq 2,  \\
m_{+} &=& m_{q \xi}, \xi \geq 2, q \geq 2, q \neq \xi.
\end{eqnarray}
As shown in the above equations, the order parameters are divided into three groups (Fig. \ref{fig:FM_mag_symmetry}).
Note that $\{ m_{q \xi} | q = \xi \}$ is not defined because $\{ \sigma_{\xi i} | i \in V_{\xi} \}$ is removed.
Figure \ref{fig:FM_rho1=0.26_mag} shows the dependence of $m_{0}$ and $m_{\pm}$ on $\Gamma$ for $\rho_{1} = 0.26$. Fig. \ref{fig:FM_rho1=0.6_mag} displays that for $\rho_{1} = 0.60$.
For $\rho_{1} = 0.26$, a metastable state appears; the first-order phase transition occurs as $\Gamma$ decreases;
$m_{0}$ and $m_{\pm}$ continuously change for $\rho_{1} = 0.60$.
The results indicate that the first-order phase transition is successfully avoided if the overlap between the candidate solution and one of the ground states exceeds a threshold.

\begin{figure}
\centering
\includegraphics[width=0.8\columnwidth]{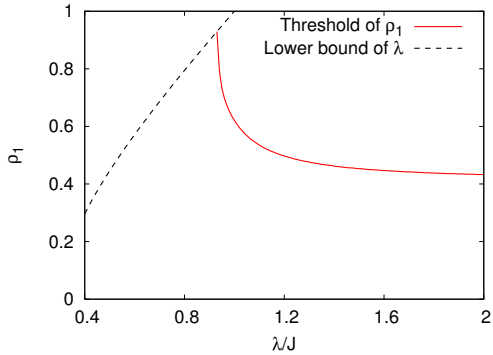}
\caption{Condition for avoiding the first-order phase transition on the $\lambda - \rho_{1}$ plane for $Q=4$.}
\label{fig:FM_prop_phase_diagram}
\end{figure}

Fig. \ref{fig:FM_prop_phase_diagram} shows the condition of avoiding the first-order phase transition on the $\lambda - \rho_{1}$ plane for $Q=4$.
The solid line represents the $\rho_{1}$ threshold. The dashed line is the lower bound of $\lambda$ to correctly encode the ground states of the FC-FM Potts model as those of the corresponding Ising model [Eq. (\ref{eq:FM_prop_problem_Ham})].
The $\rho_{1}$ threshold strongly depends on $\lambda$ and decreases as $\lambda$ is increased.
This is caused by the following reason:
the effect of the biased sampling of the zero-hot ground state becomes stronger when $\lambda$ is increased 
because the probability of obtaining infeasible solutions, which break the constraint [Eq. (\ref{eq:formulation_prop_constraint})], decreases.
We can control the biased sampling effect by using $\lambda$.
Note that unlike RA, the dynamic adjustment of $\lambda$ is not required during TMF-QA.

\begin{figure}
\centering
\includegraphics[width=0.8\columnwidth]{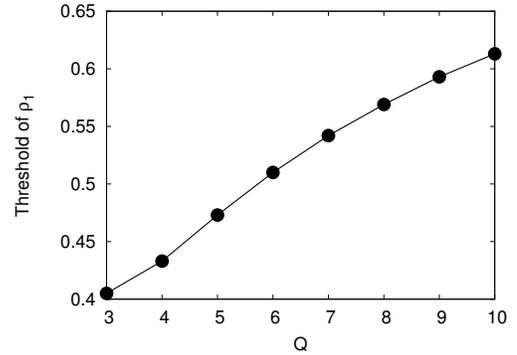}
\caption{Dependence of the $\rho_{1}$ threshold on $Q$.}
\label{fig:FM_rho1_threshold_Q_dependence}
\end{figure}

Fig. \ref{fig:FM_rho1_threshold_Q_dependence} shows the dependence of the $\rho_{1}$ threshold on $Q$.
Although the first-order phase transition can be avoided, the $\rho_{1}$ threshold becomes larger when $Q$ is increased.

\section{Conclusions}
In this study, we proposed the Ising formulation of integer optimization problems for embedding information about a candidate solution into the problem Hamiltonian.
The proposed formulation enables us to use TMF-QA to improve a candidate solution obtained by other optimization algorithms 
and combine TMF-QA with other heuristic algorithms to exploit their complementary strength.
Each heuristic algorithm escapes the local minima in different ways; hence, the hybridization of several heuristic algorithms is expected to be an effective method of obtaining a highly accurate solution for a wide range of combinatorial optimization problems.

The proposed formulation embeds information about a candidate solution by exploiting the biased sampling of the degenerated ground states in TMF-QA.
The information is embedded into the problem Hamiltonian. The number of control parameters does not increase from the original form.
In addition, the number of Ising spins decreases from the conventional one-hot encoding because the zero-hot configuration becomes one of the feasible solutions in the proposed formulation.
Therefore, the proposed formulation can be applied to QA implemented by the current version of the D-Wave quantum annealer. In addition, the hardware resource of the annealer can be efficiently exhausted.

In this work, we analytically investigated the phase transition order during TMF-QA of the FC-FM Potts model.
The results confirmed that the first-order phase transition can successfully be avoided using the proposed formulation if the overlap between the candidate solution and one of the ground states exceeds a threshold.
By applying the proposed formulation, we can use TMF-QA to improve a candidate solution.
The proposed formulation is expected to be of use in a wide range of integer optimization problems for embedding information about a candidate solution.
Confirming the validity of this expectation will be a future work.

\begin{acknowledgments}
The authors are considerably grateful to Tadashi Kadowaki and Masamichi J. Miyama for their fruitful discussions.
This work was financially supported by JSPS KAKENHI Grant No. 19H01095,  20H02168 and the Next Generation High-Performance Computing Infrastructures and Applications R{\&}D Program of MEXT, and by MEXT-Quantum Leap Flagship Program Grant Number JPMXS0120352009.
\end{acknowledgments}

\appendix
\section{Free energy of the FC-FM Potts model with the proposed Ising formulation}
We derive the free energy [Eqs. (\ref{eq:FM_prop_free_energy}), (\ref{eq:FM_prop_Ham_eff_def}), and (\ref{eq:FM_prop_h_eff_def})] of the FC-FM Potts model using the proposed Ising formulation.

The partition function is given as folllows:
\begin{equation}
Z = \mathrm{Tr} e^{- \beta \hat{\mathcal{H}}_{0}(\hat{\bm{\sigma}}^{z}) - \beta \hat{\mathcal{H}}_{q}(\hat{\bm{\sigma}}^{x})},
\end{equation}
where $\hat{\mathcal{H}}_{0}(\hat{\bm{\sigma}}^{z})$ and $\hat{\mathcal{H}}_{q}(\hat{\bm{\sigma}}^{x})$ are defined in Eqs. (\ref{eq:FM_prop_problem_Ham}) and (\ref{eq:FM_prop_driver_Ham}), respectively.
Using the Suzuki--Trotter formula yields
\begin{equation}
Z = \lim_{K \to \infty} \mathrm{Tr} \left[ \exp \left( - \frac{\beta}{K} \hat{\mathcal{H}}_{0}(\hat{\bm{\sigma}}^{z}) \right) \exp \left( - \frac{\beta}{K} \hat{\mathcal{H}}_{q}(\hat{\bm{\sigma}}^{x}) \right) \right]^{K}.
\end{equation}
By introducing the closure relations
\begin{equation}
\hat{1}(\kappa) = \sum_{\bm{\sigma}^{z}(\kappa)} \ket{\bm{\sigma}^{z}(\kappa)} \bra{\bm{\sigma}^{z}(\kappa)},  \label{eq:app_closure_relation}
\end{equation}
the partition function is rewritten as
\begin{eqnarray}
Z &=& \lim_{K \to \infty} \sum_{\bm{\sigma}^{z}(1)} \cdots \sum_{\bm{\sigma}^{z}(K)} \prod_{\kappa =1}^{K} \exp \left( - \frac{\beta}{K} \mathcal{H}_{0} \bigl( \bm{\sigma}^{z}(\kappa) \bigr) \right)  \nonumber  \\
&\times& \Braket{ \bm{\sigma}^{z}(\kappa) | \exp \left( - \frac{\beta}{K} \hat{\mathcal{H}}_{q}(\hat{\bm{\sigma}}^{x}) \right) | \bm{\sigma}^{z}(\kappa +1) },  \label{eq:app_Z_ST_formula}
\end{eqnarray}
where $\kappa$ represents the Trotter slice; $\bm{\sigma}^{z}(\kappa)$ denotes the Ising spins belonging to the $\kappa$-th Trotter slice; and $\sum_{\bm{\sigma}^{z}(\kappa)}$ denotes the summation over all the spin configurations of $\bm{\sigma}^{z}(\kappa)$.
By using Dirac's delta function and its Fourier transformation, we introduce herein the ferromagnetic order parameters and linearize the first term of $\mathcal{H}_{0}(\bm{\sigma}^{z})$ as follows:
\begin{widetext}
\begin{eqnarray}
&&\exp \left[ \frac{\beta N}{2 K} \sum_{\xi, \eta} \rho_{\xi} \rho_{\eta} \sum_{q \neq \xi} \sum_{q' \neq \eta} \tilde{J}_{\xi \eta}(q,q') \left( \frac{1}{N_{\xi}} \sum_{i \in V_{\xi}} \sigma^{z}_{qi}(\kappa) \right) \left( \frac{1}{N_{\eta}} \sum_{j \in V_{\eta}} \sigma^{z}_{q'j}(\kappa) \right) \right]  \nonumber  \\
= &&\int d \bm{m}(\kappa) \exp \left( \frac{\beta N}{2K} \sum_{\xi, \eta} \rho_{\xi} \rho_{\eta} \sum_{q \neq \xi} \sum_{q' \neq \eta} \tilde{J}_{\xi \eta}(q,q') m_{q \xi}(\kappa) m_{q' \eta}(\kappa) \right) \prod_{\xi} \prod_{q \neq \xi} \delta \left( m_{q \xi}(\kappa) - \frac{1}{N_{\xi}} \sum_{i \in V_{\xi}} \sigma^{z}_{qi}(\kappa) \right)  \nonumber  \\
\propto &&\int d \bm{m}(\kappa) \int d \tilde{\bm{m}}(\kappa) \exp \left( \frac{\beta J}{K} \sum_{\xi} \sum_{q \neq \xi} \tilde{m}_{q \xi}(\kappa) \sum_{i \in V_{\xi}} \sigma^{z}_{qi}(\kappa) \right)  \nonumber  \\
\times &&\exp \left[ \frac{\beta N}{K} \left( \frac{1}{2} \sum_{\xi, \eta} \rho_{\xi} \rho_{\eta} \sum_{q \neq \xi} \sum_{q' \neq \eta} \tilde{J}_{\xi \eta}(q,q') m_{q \xi}(\kappa) m_{q' \eta}(\kappa) - J \sum_{\xi} \sum_{q \neq \xi} \rho_{\xi} \tilde{m}_{q \xi}(\kappa) m_{q \xi}(\kappa) \right) \right],  \label{eq:app_H0_1st_term_linearize}
\end{eqnarray}
where $m_{q \xi}(\kappa)$ is the ferromagnetic order parameter of $\{ \sigma^{z}_{qi}(\kappa) | i \in V_{\xi} \}$, and $\tilde{m}_{q \xi}(\kappa)$ is the conjugate variable of $m_{q \xi}(\kappa)$, and 
\begin{eqnarray}
d \bm{m}(\kappa) &\equiv& \prod_{\xi} \prod_{q \neq \xi} dm_{q \xi}(\kappa),  \\
d \tilde{\bm{m}}(\kappa) &\equiv& \prod_{\xi} \prod_{q \neq \xi} d\tilde{m_{q \xi}}(\kappa).
\end{eqnarray}
We can rewrite $\mathcal{H}_{0}(\bm{\sigma}^{z}(\kappa))$ as follows by using Eq. (\ref{eq:app_H0_1st_term_linearize}):
\begin{eqnarray}
\exp&& \left( - \frac{\beta}{K} \mathcal{H}_{0} \bigl( \bm{\sigma}^{z}(\kappa) \bigr) \right) \propto \int d \bm{m}(\kappa) \int d \tilde{\bm{m}}(\kappa) \exp \left( - \frac{\beta}{K} \mathcal{H}^{(\mathrm{eff})}_{0} \bigl(\bm{\sigma}^{z}(\kappa) \bigr) \right)  \nonumber  \\
&&\times \exp \left[ \frac{\beta N}{K} \left( \frac{1}{2} \sum_{\xi, \eta} \rho_{\xi} \rho_{\eta} \sum_{q \neq \xi} \sum_{q' \neq \eta} \tilde{J}_{\xi \eta}(q,q') m_{q \xi}(\kappa) m_{q' \eta}(\kappa) - J \sum_{\xi} \sum_{q \neq \xi} \rho_{\xi} \tilde{m}_{q \xi}(\kappa) m_{q \xi}(\kappa) \right) \right],  \label{eq:app_H0_linearize}
\end{eqnarray}
where
\begin{equation}
\mathcal{H}^{(\mathrm{eff})}_{0} \bigl( \bm{\sigma}^{z}(\kappa) \bigr) = \frac{\lambda}{2} \sum_{\xi} \sum_{i \in V_{\xi}} \left( \sum_{q \neq \xi} \sigma^{z}_{qi}(\kappa) \right)^{2} - \sum_{\xi} \sum_{i \in V_{\xi}} \sum_{q \neq \xi} \left[ J \tilde{m}_{q \xi}(\kappa) + (Q-2) \bigl( \lambda + \tilde{h}_{\xi}(q) \bigl) \right] \sigma^{z}_{qi}(\kappa).  \label{eq:app_H0_linearize_Heff}
\end{equation}
Substituting Eqs. (\ref{eq:app_H0_linearize}) and (\ref{eq:app_H0_linearize_Heff}) into Eq. (\ref{eq:app_Z_ST_formula}), followed by the inverse operation of the closure relation [Eq. (\ref{eq:app_closure_relation})], we obtain
\begin{eqnarray}
Z &\propto&  \lim_{K \to \infty} \int d \bm{m} \int d \tilde{\bm{m}}  \nonumber  \\
&\times& \exp \left[ \frac{\beta N}{K} \left( \frac{1}{2} \sum_{\kappa} \sum_{\xi, \eta} \rho_{\xi} \rho_{\eta} \sum_{q \neq \xi} \sum_{q' \neq \eta} \tilde{J}_{\xi \eta}(q,q') m_{q \xi}(\kappa) m_{q' \eta}(\kappa) - J \sum_{\kappa} \sum_{\xi} \sum_{q \neq \xi} \rho_{\xi} \tilde{m}_{q \xi}(\kappa) m_{q \xi}(\kappa) \right) \right]  \nonumber  \\
&\times& \mathrm{Tr} \prod_{\kappa} \left[ \exp \left( - \frac{\beta}{K} \hat{\mathcal{H}}^{(\mathrm{eff})}_{0} \bigl( \hat{\bm{\sigma}}^{z} \bigr) \right) \exp \left( - \frac{\beta}{K} \hat{\mathcal{H}}_{\mathrm{q}}(\hat{\bm{\sigma}}^{x}) \right) \right],
\end{eqnarray}
where
\begin{equation}
d \bm{m} \equiv \prod_{\kappa} \prod_{\xi} \prod_{q \neq \xi} dm_{q \xi}(\kappa),
\end{equation}
and
\begin{equation}
d \tilde{\bm{m}} \equiv \prod_{\kappa} \prod_{\xi} \prod_{q \neq \xi} d \tilde{m}_{q \xi}(\kappa).
\end{equation}
We then inversely operate the Suzuki--Trotter formula as follows after applying the static approximation:
\begin{eqnarray}
m_{q \xi} &\equiv& m_{q \xi}(\kappa),  \\
\tilde{m}_{q \xi} &\equiv& \tilde{m}_{q \xi}(\kappa).
\end{eqnarray}
The resulting expression of the partition function is given as follows:
\begin{equation}
Z = \int d \bm{m} \int d \tilde{\bm{m}} e^{- \beta N f(\bm{m}, \tilde{\bm{m}})},
\end{equation}
where
\begin{equation}
f(\bm{m}, \tilde{\bm{m}}) = - \frac{1}{2} \sum_{\xi, \eta} \rho_{\xi} \rho_{\eta} \sum_{q \neq \xi} \sum_{q' \neq \eta} \tilde{J}_{\xi \eta}(q,q') m_{q \xi} m_{q' \eta} + J \sum_{\xi} \sum_{q \neq \xi} \rho_{\xi} \tilde{m}_{q \xi} m_{q \xi} - \frac{1}{\beta} \sum_{\xi} \rho_{\xi} \log \mathrm{Tr} e^{- \beta \hat{\mathcal{H}}^{(\mathrm{eff})}_{\xi}},  \label{eq:app_FM_prop_free_energy}
\end{equation}
\begin{equation}
\hat{\mathcal{H}}^{(\mathrm{eff})}_{\xi} = \frac{\lambda}{2} \left( \sum_{q \neq \xi} \hat{\sigma}^{z}_{q} \right)^{2} - \sum_{q \neq \xi} \tilde{h}^{(\mathrm{eff})}_{\xi}(q) \hat{\sigma}^{z}_{q} - \Gamma \sum_{q \neq \xi} \hat{\sigma}^{x}_{q},  \label{eq:app_FM_prop_Ham_eff_def}
\end{equation}
and
\begin{equation}
\tilde{h}^{(\mathrm{eff})}_{\xi}(q) = J \tilde{m}_{q \xi} + (Q-2) \bigl[ \lambda + J ( \rho_{q} - \rho_{\xi} ) \bigr].  \label{eq:app_FM_prop_h_eff_def}
\end{equation}
Eqs. (\ref{eq:app_FM_prop_free_energy}), (\ref{eq:app_FM_prop_Ham_eff_def}), and (\ref{eq:app_FM_prop_h_eff_def}) are identical to Eqs. (\ref{eq:FM_prop_free_energy}), (\ref{eq:FM_prop_Ham_eff_def}), and (\ref{eq:FM_prop_h_eff_def}), respectively.
\end{widetext}

\bibliography{reference}
\end{document}